\documentclass[12pt]{article}
\usepackage{graphicx}
\begin{document}
\begin{titlepage}
\begin{center}

\begin{large}
\textbf{Q-PET: PET with 3rd Eye}

Quantum Entanglement based Positron Emission Tomography
\end{large}

\end{center}
\end{titlepage}
\begin{quote}
\begin{center}

\textbf{Q-PET: PET with 3rd Eye}

\textbf{Quantum Entanglement Based Positron Emission Tomography}
\end{center}
\end{quote}

\begin{center}
\textit{Sunil Kumar\footnote{corresponding author}, Sushil Singh Chauhan, Vipin Bhatnagar}
\end{center}
\begin{center}
\textit{corresponding e-mail : saroha.sk@gmail.com}
\end{center}

\begin{abstract}
In the present ongoing study, we are proposing a prototype model for positron emission tomography detection technology by introduction of a new discriminatory window parameter. It can be a new generation PET detection technique. We introduced Polarization Measurement [2] of the annihilation photons(generated from the annihilation of positron and electron) as an additional parameter in proposed prototype, to correlate annihilation photons of a particular annihilation event. The motivation behind this introduction is \textit{Quantum Entanglement} relation between the two annihilation photons. These two oppositely emitted photons are linearly polarized at right angle to each other [3]. Simulations studies for this research work are undergoing and some preliminary results are presented here.
\end{abstract}
\textbf{keywords} : quantum entanglement, annihilation, polarization, PET
\section{Introduction}

Medical imaging is the field in which radiation is used for imaging the body of the diseased patient. For this purpose, many such systems has been developed to serve the man kind. X-Ray radiography, Computed Tomography (CT), Ultrasonography, Magnetic Resonance Imaging,  Nuclear Medicine Imaging and Positron Emission Tomography (PET) are some techniques generally used in medical imaging.
Out of these, PET is widely used for the staging, restaging, drug and therapy response of the patients diagnosed with cancer. 

PET enables us to get morphological/functional imaging of bio-distribution of positron emitting radionuclide or radiopharmaceutical introduced purposely within the body of the patient/animal. Current PET detection technique involves co-incidence detection technique to correlate the two annihilation photons emitted in the almost exactly opposite direction detected by a ring type scintillation-based detection system [1]. 
 
PET involves a technique to get information of position of annihilation event of positronium decaying to number of photons (can decay to more than 2 photon). Position information of the events helps in constrcuting an image of the distribution of positron emitting radionuclide in the body of patient[1].

To reconstruct the raw data in an informative PET image, one needs to investigate exactly the true events (paired photons) from the random, scattred and multiple events [2].

 False coincidences introduce noise and contrast lost in the reconstructed image and also enhance the chances to misinterpretation. PET systems involve the two windows for the selection of true events which are enegy window and timing window. At present, time and energy windowing, which are applied over the primary coincidence data, discard all multiple events, as well as a considerable fraction of unscattered coincidence events due to the relatively poor energy resolution of current detectors [2].

Besides these parameters for the selection of true coincidence, there is an other parameter which, possiblly can accuratly measure the true coincidence. And that parameters enables us to introduced another selection window (which we are calling as the 3rd EYE) based on quantum entanglement of the two annihilation photons emitted by para-positronium.


\section{Motivation}

Although conventional PET system are working fine with the current technology of event detection using collinearity for reconstruction, but the technique of detection introduces uncertainity in position due to non-collinearity of the two emitted photon. Using gaseous detector for tracking of recoil electron, provides direction of entrance of photon within some uncertainity in solid angle $d\Omega$.
The motivation towards this work comes from the fact that using Compton scattering, we can find the polarization of the photons and that can be used to identify the correlated true annihilation photons.

Two photons emitted are linearly polarized such that the polarization vectors are orthogonal to each other i.e. having a quantum entangled state in which both the photons have their planes of polarization perpendicular to each other[3]. This quantum entangled state can work as a discrimantory window for the identification of true annihilation events after clearing the first two windows. So if one can measure the the polarization of each photon and can identify the orthogonal relation between them, this technique can work as a powerful tool to identify accurately the two annihilation photons. 


\section{Physics}

\subsection{Positron Annihilation}

Electron and positron can form a bound state exactly like electron and proton. That bound state is called positronium and is purely leptonic object. 
The ground state of positronium, like that of hydrogen atom, has two possible configurations depending on the total spin of the electron and the positron. If total spin of positronium is S = 0 then that state is called para-positronium (p-Ps) with a mean lifetime 0.125 ns. Positronium with parallel oriented spins (with total spin S = 1) is known as ortho-positronium (o-Ps) and it has mean lifetime  142 ns [11].
Positronium, due to conservation of charge conjugation symmetry, can decay into even (p-Ps) or odd (o-Ps) number of gamma particles. However decays into four or more gammas are negligible - for para positronium decay into four photons have branching ratio $1.44 x 10_{-6}$ [10].
\subsection{Intercation Mechanism of Gamma Rays}
There are three major interaction mechanisms of photons with matter and three of them are the means by which photons are detected. The three mechanism are:
1. Photoelectric absorption
2. Compton scattering
3. Pair production

The predominant mode of interaction depends on the energy of the incident photons and the atomic number of the material with which they are interacting. At low energies, in high atomic number materials, the photoelectric effect is the main interaction of photons with the material. At intermediate energies in low atomic number materials the dominant interaction is Compton scattering. At very high energies, the main mechanism by which photons are detected is pair production.
It should be noted that photons deposit energy in a material by transferring their energy to a secondary charged particle such as electron, and it is the energy imparted to the electron that is deposited in the detector.

\subsubsection{Photoelectric Effect}

In this case an incident photon ejects an electron, designated a photoelectron, from an absorber atom shell and the photon is completely absorbed. The kinetic energy of the photoelectron E{e} is given by

$E_{e} = E_{\gamma} - E_{BE}$

where $E_{\gamma}$ is the incident photon energy and $E_{BE}$ the electron binding energy. $E_{BE}$ is usually small compared to $E_{\gamma}$, so that the photoelectron carries off most of the photon energy.

The vacancy in the electron shell is quickly filled by electron rearrangement, electrons from higher energy levels fill the vacancy. The binding energy is liberated as characteristic (fluorescence) X-rays or Auger electrons. Assuming the binding energy is absorbed, the feature that appears in the measured spectrum as a result of photoelectric events is a full energy peak.
The interaction cross section of the photoelectric absorption [8] has a dependence of
\begin{center}
$\tau =\frac{Z^{5}}{{E_{\gamma}}^{3.5}}$
            (Z: atomic number)
\end{center}

\subsubsection{Compton scattering}

In Compton Scattering, single electron works as a target and the recoil energy of the electron is larger than the electron binding energy. Therefore, the incident photon transfers part of the energy and the recoil electron is emitted after scattering [8].

\section{Method}

Compton cross section depends on the photon polarization and same dependance make us able to calculate polarisation of the photons involved in the annihilation process[9].It is also implied by Quantum Electrodynamics that the two photons emitted in an electron positron annihilation process are polarized orthogonal(perpendicular) to each other[9].

In this study we have planned for using the angular corelation between compton scattered annihilation photons. For this purpose Klein-Nishina Model for Compton Scattering will be used as a theoretical basis.

Geant4 Simulation will provide the requisite variables for the calculations of the angular correlation.

The differential cross-section of Compton scattering [7] is expressed as

\begin{center}
\textbf{{\large $\frac{d\sigma }{d\Omega } = \frac {r_0^2}{2}\left ( \frac{E'}{E} \right )^2 \left ( \frac{E}{E'} + \frac{E'}{E}-2\sin^2\theta \cos^2\eta  \right) $}}
\end{center}

where 

$r_0$ = classical electron radius

E   = energy of incident gamma ray

E'  = energy of scattered gamma ray 

$\theta$  = angle of scattering​

$\eta$  = angle between plane of scattering
\& plane of polarization​

\begin{center}
\includegraphics[scale=0.5]{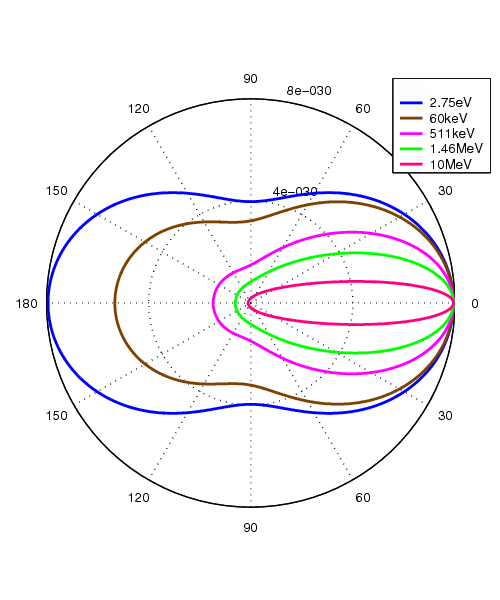}
\end{center}
\begin{center}
Fig 7. Klein-Nishina Differential Cross Section for different energies
\end{center}

\section{Prototype}
\subsection{Geometry}

Q-PET consist of 16 units of detector arranged in a ring of inner radius 30.78 cm and outer radius 47.04 cm. Each unit consist of one gaseous detector (Scatterer/Tracker) and one scintillation block.

\begin{center}
\includegraphics[scale=0.6]{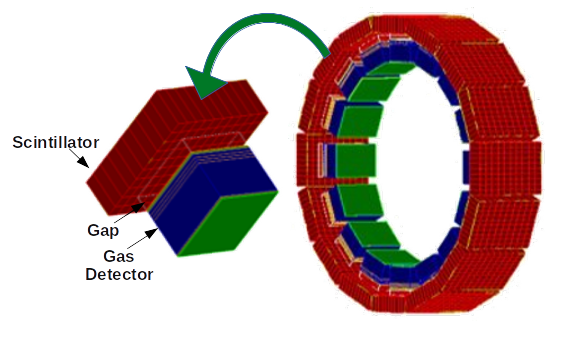}
\end{center}
\begin{center}
A Geant4 constructed geometry
\end{center}

\textbf{Tracker/Scatterer}

In Tracker Section, Gas Chamber has been introduced for the tracking purposes. Material used for the gas chamber is $XeCO_{2}C_{4}F_{10}$ in  proportion ${Xe:CO_{2}:C_{4}F_{10}::50:15:35}$. The material($XeCO_{2}C_{4}F_{10}$) so chosen for the gas detector contains Xenon, which is the highest density Noble gas can be used for the proportional chambers. To increase the probability of 511 keV photons interaction with gas molecule, density of the gas should be large. Besides this one has to consider some factor i.e. interaction cross-section for the desired process, density should be enough to have the interaction probability to get the compton interaction. 

Gas detector has dimension \textit{10 cm x 10 cm x 8.96 cm}. Gaseous detector will provide the position of very first interction of annihilation photon in the gaseous medium. Signal generated by the gaseous detector also give information ofthe track of recoil electron. Time Projection Chamber (TPC) configuration can be used to get the third co-ordinate (depth of interction) point. Besides that some other methods are there to get the depth of interaction point.

Tracking of recoil electron can provide the additional benefit of getting the real direction of gamma-ray entrance within some uncertainity. This feature is very important for large diameter scanners.

\textbf{Scintillator}

In Scintillator Section, $CdWO_{4}$ crystals are used to absorb the scattered gamma ray completely. Scintillation Block of each unit have 10 x 10 arrangement of crystals of dimension \textit{1.58 cm x 1.58 cm x 5 cm} each.  
This section will provide position and energy of the scattered gamma ray.

\section{Summary}

The polarization measurement technique can work as a discriminatory window for the identification of annihilation photons. The present study suggest probability of photon interaction in the gaseous detection ~0.02.
Work is going on to improve the interaction probability and to reconstruct the image from the raw data. The results will be updated soon.

\section{References}

[1]PET Physics, Instrumentation and Scanners by MIchael E Phelps.

[2]Polarisation-based coincidence event discrimination: an in silico study towards a feasible scheme for Compton-PET by M Toghyani, J E Gillam, A L McNamara and Z Kuncic.

[3]The Atomic Nucleus by by Robley D. Evans TATA McGRAW-HILL Publishing Company LTD. Bombay - New Delhi.

[4]Electron, Positron and Photon Polarimetry by Johannes Marinus Hoogduin.

[5]J-PET: a new technology for the whole-body PET imaging by S. Niedźwiecki \& P. Białas et al.

[6]Performance of a New Electron-Tracking Compton Camera under Intense Radiations from a Water Target irradiated with a Proton Beam by Y. Matsuoka \& T. Tanimori et al.

[7]Polarimetry at high energies by Wojtek Hajdas and Estela Suarez-Garcia.

[8]Radiation Detection \& Measurement by Glenn F Knoll.

[9]The Angular Correlation of Polarization of Annihilation Radiations by Nicholas Carlin and Peter Woit.

[10]Precision Study of Positronium: Testing Bound State QED Theory by Savely G. Karshenboim.

[11]First test of O$(\alpha^2)$ correction of the orthopositronium decay rate by . Kataoka, S. Asai and T. Kobayash.
\end{document}